\documentclass[aps,pre,eqsecnum,showpacs,draft]{revtex4}
\usepackage{amsmath,bm}

\begin{document}

\title{Disclination Motion in Hexatic and Smectic-$C$ Films}

\author{E. I. Kats $^{1,2}$, V. V. Lebedev $^{1,3}$,
and S. V. Malinin $^{1,4}$}
\affiliation{$^1$ L. D. Landau Institute for Theoretical Physics, RAS \\
117940, Kosygina 2, Moscow, Russia; \\
$^2$ Laue-Langevin Institute, F-38042, Grenoble, France; \\
$^3$ Theoretical Division, LANL, Los-Alamos, NM 87545, USA \\
$^4$ Forschungszentrum J\"ulich, D-52425, J\"ulich, Germany.}

\date{\today}

\begin{abstract}

  We present theoretical study of a single disclination motion in a
  thin free standing hexatic (or smectic-$C$) film, driven by a
  large-scale inhomogeneity in the bond (or director) angle. Back-flow
  effects and own dynamics of the bond angle are included. We find the
  angle and hydrodynamic velocity distributions around the moving
  disclination, what allows us to relate the disclination velocity to
  the angle gradient far from the disclination. Different cases are
  examined depending on the ratio of the rotational and shear
  viscosity coefficients.

\end{abstract}


\maketitle

\section*{Introduction}

Physics of thin liquid-crystalline films has been a recurrent hot
topic during the past decade because of their intriguing physical
properties and wide applications in display devices, sensors, and
for many other purposes. Hexatic and smectic-$C$ liquid crystalline
films, we consider, belong to two-dimensional systems with
spontaneously broken continuous symmetry. Therefore an essential
role in the behavior of the films is played by vortex-like
excitations (disclinations). Defects are almost necessarily present
in liquid crystals, and their dynamics plays a crucial role in the
overall pattern organization. Early studies of defects focused on
classifying the static properties of the defects and their
interactions \cite{gen,chandra}. More recently the focus has moved
to examining the dynamics of defects (see, e.g., \cite{OP00} and
references herein). Note that though in most practical applications
of liquid crystals, such as traditional display devices, defects
destroy an optical adjustment, and therefore are not desirable,
there are novel display designs (bistable, multidomain liquid
crystalline structures) just exploiting defect properties.

Unfortunately, theoretical researches of dynamic characteristics of
the films are in a rather primitive stage, in spite of the fact
that experimental dynamic studies are likely to be more fruitful.
It is largely accounted for by a complexity of dynamic phenomena in
the films, and a complete and unifying description of the problem
is still not available. Moreover, some papers devoted to these
problem (dynamics of defects) claim contradicting results. These
contradictions come mainly from the fact that different authors
take into account different microscopic dissipation mechanisms, but
partially the source of controversy is related to semantics, since
different definitions for the forces acting on defects are used
(see, e.g., discussion in \cite{SO97}).

In this paper we examine theoretically the disclination dynamics in
free standing hexatic and smectic-$C$ films at scales, much larger
than their thickness, where the films can be treated as $2d$
objects. Our investigation is devoted to the first (but compulsory)
step of defect dynamics studies: a single point disclination in a
hexatic or smectic-$C$ film. A number of theoretical efforts
\cite{PL88,BP88,PR90,RK91,DE96} deal with similar problems. Our
justification for adding one more paper to the topic is the fact
that we did not see in the literature a full investigation of the
problem taking into account hydrodynamic back-flow effects.
Evidently, these effects can drastically modify dynamics of
defects. The goal of this work is to study the disclination motion
in free standing liquid crystalline films on the basis of
hydrodynamic equations, containing some phenomenological parameters
(elasticity modulus, shear and rotational viscosity coefficients).

In our approach the disclination is assumed to be driven by a
large-scale inhomogeneity in the bond or director angle providing a
velocity for the disclination motion (relative to the film). As a
physical realization of such non-uniform angle field, one can have
in mind a system of disclinations distributed with a finite
density. Then the inhomogeneity near a given disclination is
produced by fields of other disclinations. One can also think about
a pair of disclinations of opposite topological charges, then this
inhomogeneity is related to mutual orientational distortion fields
created by each disclination at the point of its counterpart. In
fact, the majority of experimental and numerical studies of
disclination motions in liquid crystals and their models
\cite{CP80,CS87,PT91,YP93,LD87,DO00,DO01,TD02,FU98} is devoted just
to investigation of the dynamics of two oppositely charged defects.
We solve the hydrodynamic equations and find bond angle and flow
velocity distributions around the moving disclination. The results
enable us to relate the disclination velocity and a gradient of the
bond angle far from the moving disclination.

An obvious object, where our results can be applied, is the film
dynamics near the Berezinskii-Kosterlitz-Thouless phase transition.
The static properties of the films near the transition have been
investigated in a huge number of papers starting from the famous
papers by Berezinskii \cite{BE70} and Kosterlitz and Thouless
\cite{KT73}. There is some literature discussing the theory of
dynamic phenomena associated with vortex-like excitations in
condensed matter physics: vortices in type-II superconductors (see,
e.g., \cite{BF94}), vortices in superfluid $^4He$ and $^3He$ (see,
e.g., \cite{HV56,SO87}), dislocations in crystals and disclinations
(and other topological defects) in liquid crystals (see
\cite{WC85,DP86,rev,CP80,CS87,PT91,MC92,YP93,LD87}). However, most
of the theoretical works on the topic start from phenomenological
equations of motion of the defects, and our aim is to derive the
equations and to check their validity.

Our paper has the following structure. Section \ref{basiceq}
contains basic hydrodynamic equations for liquid crystalline films
necessary for our investigation. In Section \ref{mobility} we find
the bond angle and the flow velocity around the uniformly moving
disclination in different regions, that allows us to relate the
disclination velocity to the angle gradient far from the
disclination. Different cases, depending on the ratio of the
rotational and shear viscosity coefficients, are examined in
Section \ref{cases}. The last section \ref{discussion} contains
summary and discussion. Appendix is devoted to details of
calculations of the velocity and bond angle fields around the
moving disclination. Those readers who are not very interested in
mathematical derivations can skip this Appendix finding all
essential physical results in the main text of the paper.

\section{Basic relations for hexatic and smectic-$C$ films}
\label{basiceq}

Let us formulate basic relations needed to describe disclination
motion in thin liquid crystalline films. Here we investigate freely
suspended films on scales, larger than the thickness of the films,
where they can be treated as two-dimensional objects. Such object
can be described in terms of a macroscopic approach, containing
some phenomenological parameters.

There are different types of the film ordering. Experimentally,
hexatic and smectic-$C$ films are observed. In these films, as in
$3d$ nematic liquid crystals, the rotational symmetry is
spontaneously broken. The smectic-$C$ films are characterized by
the liquid crystalline director which is tilted with respect to the
normal to the film, therefore there is a specific direction along
the film. In the hexatic films molecules are locally
arranged in a triangular lattice, however the lattice is not an
ideal one. The positional order does not extend over distances
larger than a few molecular size. Nevertheless, the bond order
extends over macroscopic distances. Therefore the phase is
characterized by the $D_{6h}$ point group symmetry. The order
parameter for the smectic-$C$ films is an irreducible tensor
$Q_{\alpha\beta}$ (the subscripts denoted by Greek letters run two
values, since we treat the films as $2d$ objects). The hexatic
order parameter is an irreducible symmetric sixth-rank tensor
$Q_{\alpha\beta\gamma\delta\mu\nu}$ \cite{gur}, both such tensors
have two independent components. Note that the order in the
smectic-$C$ films can be readily observed, by looking for in-plane
anisotropies in quantities such as the dielectric permeability
tensor. Because of its intrinsic sixfold rotational symmetry,
hexatic orientational order is hardly observable. However, it can
be detected, e.g., as a sixfold pattern of spots in the in-plane
monodomain $X$-ray structure factor, proportional to
$Q_{\alpha\beta\gamma\delta\mu\nu}$ (see, e.g., \cite{OP00} and
references herein).

In accordance with the Goldstone theorem, in the both cases of
smectics-$C$ and hexatics the only ``soft'' degree of freedom of
the order parameter is relevant on large scales, which is an angle
$\varphi$ (like the phase of the order parameter for the superfluid
$^4He$). It is convenient to express a variation of the order
parameter in terms of a variation of the angle $\varphi$. For the
smectic-$C$ films the relation is
\begin{eqnarray}
\label{gur2}
\delta Q_{\alpha\beta} =
- \delta \varphi \epsilon _{\alpha \mu } Q_{\mu\beta}
- \delta \varphi \epsilon _{\beta \mu } Q_{\alpha\mu} \,,
\end{eqnarray}
where $\epsilon_{\alpha\rho}$ is the two-dimensional antisymmetric
tensor. For the hexatic films
\begin{eqnarray}
\label{gur1}
\delta Q_{\alpha \beta \gamma \delta \mu \nu }
=-\delta\varphi\epsilon _{\alpha \rho }
Q_{\rho \beta \gamma \delta \mu \nu}+ \dots \, ,
\end{eqnarray}
where dots represent a sum of all possible combinations of the same
structure. Thus for the both films symmetries the order parameter
can be characterized by its absolute value $|Q|$ and the phase
$\varphi$ which are traditionally represented as a complex quantity
$\Psi$ (see, e.g., \cite{rev}). For the hexatic films the quantity
is written as $\Psi=|Q|\exp(6i\varphi)$, and for the smectic-$C$
films the quantity is written as $\Psi=|Q|\exp(2i\varphi)$.

The angle $\varphi$ should be included into the set of the
macroscopic variables of the films. A convenient starting point of
the consideration is the energy density (per unit area) $\rho
v^2/2+\varepsilon$, where $\rho$ is the $2d$ mass density, ${\bm
v}$ is the film velocity, and $\varepsilon$ is the internal energy
density. The latter is a function of the mass density $\rho$, the
specific entropy $\sigma$, and the angle $\varphi$. In fact,
$\varepsilon$ depends on $\nabla\varphi$, since any homogeneous
shift of the angle $\varphi$ does not touch the energy. For the
hexatic films, leading terms of the energy expansion over gradients
of $\varphi$ are
\begin{eqnarray}
\label{aa1}
\varepsilon = \varepsilon_0 (\rho,\sigma) +
\frac{K}{2}(\nabla\varphi)^2 ,
\end{eqnarray}
where $K$ is the only (due to hexagonal symmetry) orientational
elastic module of the film. For smectic-$C$ films one should
introduce two orientational elastic modules, longitudinal and
transversal with respect to the director $\bm n$, determining a
specific direction in the film:
$Q_{\alpha\beta}\propto\delta_{\alpha\beta}-2n_\alpha n_\beta$.
However, the anisotropy in real smectic-$C$ films is weak because
of the small tilt angle of the director. Moreover, fluctuations of
the director lead to a renormalization of the modules, and on large
scales there occurs an isotropization of a smectic-$C$ film
\cite{NP79}. Therefore in this case we can use the same expression
(\ref{aa1}) for the elastic energy.

The complete dynamic equations for the freely suspended liquid
crystalline films, valid on scales larger than their thickness, can
be found in \cite{KL93}. We are going to treat a quasistationary
motion of the disclination. At such motions hard degrees of freedom
are not excited. By other words, we can accept incompressibility
and neglect bending deformations which are suppressed by the
presence of the surface tension in freely suspended films. In the
same manner, for the quasistationary disclination motion the
thermo-diffusive mode is not excited, what implies the isothermal
condition. For freely suspended films such effects as substrate
friction (relevant, say, for Langmuir films) are absent. Thus at
treating the disclination motion we can consider the system of
equations for the velocity ${\bm v}$ and the angle $\varphi$
solely. The equations are formulated at the conditions $\rho={\rm
const}$, $T={\rm const}$ (where $T$ is temperature) and $\nabla{\bm
v}=0$.

The equation for the velocity follows from the momentum density
${\bm j}=\rho{\bm v}$ conservation law
\begin{eqnarray}
\partial_t j_\alpha=-\nabla_\beta\left[
T_{\alpha \beta }-\eta(\nabla_\alpha v_\beta
+ \nabla_\beta v_\alpha)\right] \,,
\label{lli1} \end{eqnarray}
where $T_{\alpha\beta}$ is the reactive (non-dissipative) stress
tensor and $\eta$ is the $2d$ shear viscosity coefficient of the
film. For the two-dimensional hexatics the reactive stress tensor
is (see \cite{KL93}, chapter 6)
\begin{eqnarray} &&
T_{\alpha\beta}=\rho v_\beta v_\alpha
- \varsigma \delta_{\alpha \beta }
+K\nabla_\alpha \varphi \nabla_\beta \varphi
-\frac{K}{2} \epsilon_{\alpha\gamma }
\nabla _\gamma \nabla_\beta \varphi
-\frac{K}{2} \epsilon_{\beta \gamma}
\nabla_\gamma \nabla_\alpha\varphi \,,
\label{a7} \end{eqnarray}
where $\varsigma=\varepsilon-\rho\partial\varepsilon/\partial\rho$
is the surface tension. Note that the ratio $K\rho/\eta^2$ is a
dimensionless parameter which can be estimated by substituting
$3d$ quantities instead of $2d$ ones (since all the $2d$ quantities
can be estimated as corresponding $3d$ quantities multiplied by the
film thickness, and the latter drops from the ratio). For all
known liquid crystals the ratio is $10^{-3}\div10^{-4}$ (see,
e.g., \cite{gen,chandra,BC96,OP00}), and can be, consequently,
treated as a small parameter of the theory.

The second dynamic equation, the equation for the bond angle, is
\begin{eqnarray}
\partial_t \varphi =
\frac{1}{2}\epsilon_{\alpha\beta}\nabla_\alpha v_\beta
-v_\alpha \nabla_\alpha \varphi
+ \frac{K}{\gamma} \nabla^2 \varphi \,,
\label{IIi2} \end{eqnarray}
where $\gamma$ is the so-called $2d$ rotational viscosity
coefficient. We did not find in the literature values of the
coefficient for thin liquid crystalline films. For bulk liquid
crystals (see, e.g., \cite{gen,chandra,BC96,OP00}) the $3d$
rotational viscosity coefficient is usually few times larger than
the $3d$ shear viscosity coefficient. Therefore one could expect
$\gamma>\eta$. However, to span a wide range of possibilities below
we treat $\varGamma=\gamma/\eta$ as an arbitrary parameter.

If disclinations are present in the film, it is no longer possible
to define a single-valued continuous bond-angle variable $\varphi$.
However, the order parameter is a well-defined function, which goes
to zero at a disclination position. The gradient of $\varphi(t,{\bm
r})$ is a single-valued function of ${\bm r}$ which is analytic
everywhere except for an isolated point which is the position of
the disclination. The phase acquires a certain finite increment at
each turn around the disclination
\begin{eqnarray}
\oint \mbox dr_\alpha \,\nabla_\alpha\varphi
={2\pi} s \,,
\label{a3} \end{eqnarray}
where the integration contour is a closed loop around the
disclination position, going anticlockwise, and $s$ is a
topological charge of the disclination. For hexatic ordering
$s=(1/6)n$, and for nematic orientational order $s=(1/2)n$, where
$n$ is an integer number. One can restrict oneself to the
disclinations with the unitary charge $n$ (i.e. with $s=\pm1/6$ for
the hexatic films, $s=\pm1/2$ for the smectic-$C$ films) only,
since disclinations with larger $|s|$ possess a larger energy than
the set of unitary disclinations with the same net topological
charge, and thus the defects with larger charges are unstable with
respect to dissociation to unitary ones. Therefore the
disclinations with the charges $|n|>1$ do not play an essential
role in the physics of the films \cite{gen,chandra,BC96,OP00}. To
bring expressions, given below, into a compact form, we keep the
notation $s$ for the topological charge.

The static bond angle is determined by the stationary condition
$\delta E/\delta\varphi=0$, where
\begin{eqnarray}
E=\int\mbox d^2r\,
\left(\frac{\rho}{2}v^2+\varepsilon\right)
\nonumber \end{eqnarray}
is the energy of the film.
For the energy density (\ref{aa1}) the condition is reduced to the
Laplace equation
\begin{eqnarray}
\nabla ^2 \varphi = 0  \,.
\label{a2} \end{eqnarray}
There is a symmetric solution $\varphi_0$ for an isolated
motionless disclination which satisfies Eqs. (\ref{a3},\ref{a2}),
which has the following gradient
\begin{eqnarray}
\nabla_\alpha\varphi_0
=-s\epsilon_{\alpha\beta}
\frac{r_\beta -R_\beta}{({\bm r}-{\bm R})^2}.
\label{a8} \end{eqnarray}
Here $\bm R$ is a position of the disclination. If the origin of
the reference system is placed to this point then one can write
$\varphi_0=s\arctan(y/x)$, where $x$ and $y$ are coordinates of the
observation point $\bm r$. If some dynamic processes in the film
occur, then $\varphi$ varies in time and the distribution
(\ref{a8}) is disturbed. It is perturbed also due to the presence
of the bond angle distortion related to boundaries or other
disclinations.

Below, we have in mind a case, when a system of a large number of
disclinations (with an uncompensated topological charge) is
created. For the $3d$ nematics it can be done rather easily
\cite{OP00,gen,chandra}, since due to the intrinsic elastic
anisotropy the energies of positively and negatively charged
defects are different. We do not aware of experimental or
theoretical studies of defect nucleation mechanisms in free
standing smectic or hexatic films. Hopefully, a situation with a
finite $2d$ density of defects can be realized for the films as
well (for instance, the defects could appear even spontaneously as
a mechanism to relieve frustrations in chiral smectic or hexatic
films, i.e. analogously to the formation of the Abrikosov vortex
lattice in superconductors \cite{KS01}). Examining a motion of the
disclination in this case, we investigate a vicinity of the
disclination of the order of the inter-disclination distance. Then
far from the disclination the bond angle $\varphi$ can be written
as a sum $\mbox{const}+\bm u\bm r$, where $u$ is much larger than
the inverse inter-disclination distance (since the number of
disclinations is large). Near the disclination position the bond
angle $\varphi$ can be approximated by the expression (\ref{a8}).
Our main problem is to establish a general coordinate dependence of
$\varphi$ and $\bm v$, which, particularly, enables one to relate
the bond gradient $\bm u$ and a velocity of the disclination.

\section{Flow and angle field around a uniformly moving disclination}
\label{mobility}

Here we proceed to the main subject of our consideration, which is
a single disclination, which is driven by a large-scale
inhomogeneity in the bond angle $\varphi$. A disclination velocity
is determined by an interplay of the hydrodynamic back-flow and by
the own dynamics of the angle $\varphi$. To find the velocity, one
has to solve the system of equations
(\ref{lli1},\ref{a7},\ref{IIi2}) with the constraint (\ref{a3}),
requiring a suitable asymptotic behavior. As we explained in the
previous section, at large distances from the disclination, the
angle $\varphi$ is supposed to behave as $\mbox{const}+\bm u \bm
r$. In addition, we are going to work in the reference system,
where the film as a whole is at rest. That means, that the flow
velocity, excited by the disclination, has to tend to zero far from
the disclination.

Let us consider a situation, when the disclination moves with a
constant velocity $\bm V$. In the case both the angle $\varphi$ and
the flow velocity are functions of $\bm r-\bm V t$ (where $\bm
R=\bm V t$ is the disclination position). Then the equation for the
velocity (\ref{lli1}) is written as
\begin{eqnarray} &&
\rho(V_{\beta}-v_\beta) \nabla_\beta v_\alpha
+\eta\nabla^2 v_\alpha +\frac{K}{2}
\epsilon_{\alpha\beta}\nabla_\beta \nabla^2 \varphi
-K\nabla_\alpha\varphi \nabla^2 \varphi +\nabla_\alpha
\left[\varsigma - \frac{K}{2}(\nabla\varphi)^2 \right]= 0 \,.
\label{velocb} \end{eqnarray}
One can omit the first (inertial) term in the left-hand side of
(\ref{velocb}), which is small because of the smallness of the
parameter $K\rho/\eta^2$. Then we obtain from Eqs. (\ref{lli1}-\ref{a7})
\begin{eqnarray} &&
\nabla^2 v_\alpha+\frac{K}{2\eta}
\epsilon_{\alpha\beta}\nabla_\beta \nabla^2 \varphi
-\frac{K}{\eta}\nabla_\alpha\varphi \nabla^2 \varphi
+\nabla_\alpha\varpi=0 \,,
\label{veloc} \end{eqnarray}
where $\varpi=\eta^{-1}[\varsigma-(K/2)(\nabla\varphi)^2]$. The
equation for the angle $\varphi$, following from Eq. (\ref{IIi2}),
at the same conditions is
\begin{eqnarray} &&
\nabla^2\varphi  + \frac{\gamma}{K}V_\alpha\nabla_\alpha\varphi =
\frac{\gamma}{K}v_\alpha \nabla_\alpha\varphi
-\frac{\gamma}{2K}\epsilon_{\alpha\beta}
\nabla_\alpha v_\beta \,.
\label{phiphi} \end{eqnarray}

We are looking for a solution, which is characterized by the
following asymptotic behaviour at $r\to\infty$: the velocity $\bm
v$ vanishes and $\nabla\varphi$ tends to a constant vector $\bm u$.
It is clear from the symmetry of the problem, that the gradient
$\bm u$ of the bond angle is directed along the $Y$-axis, provided
the velocity is directed along the $X$-axis. Therefore $\varphi\to
uy$ at $r\to\infty$. Our problem is to find a relation between the
quantities $V$ and $u$, that is between the disclination velocity
and the bond angle gradient far from the disclination. There are
two different regions: large distances $r\gg u^{-1}$ and the region
near the disclination $r\ll u^{-1}$. At large distances corrections
to the leading behavior $\varphi\approx uy$ are small and the
problem can be treated in the linear over these corrections
approximation. In the region near the disclination $\varphi$ is
close to the static value (\ref{a8}) and the flow velocity $\bm v$
is close to the disclination velocity ${\bm V}$ (the special case
when the ratio $\gamma/\eta$ is extremely small will be discussed
in Subsection \ref{sub:extr}). Those two regions are separately
examined below. The relation between $u$ and $V$ can be found by
matching the asymptotics at $r\sim u^{-1}$. As a result, one
obtains
\begin{eqnarray} &&
V=\frac{K}{\eta}C u \,,
\label{final} \end{eqnarray}
where $C$ is a dimensionless factor depending on the dimensionless
ratio $\varGamma=\gamma/\eta$. This factor $C$ is of the order of
unity if $\varGamma\sim1$. We are interested in the asymptotic
behavior of $C$ at small and large $\varGamma$.

\subsection{Region near the disclination}
\label{sub:near}

Let us consider the region $r\ll u^{-1}$. Here one can write
\begin{eqnarray} &&
\varphi=\varphi_0(\bm r-\bm R)
+\varphi_1(\bm r-\bm R) \,,
\label{rg1} \end{eqnarray}
where $\bm R=\bm V t$ is the disclination position, the angle
$\varphi_0$ corresponds to an equilibrium bond angle around an
unmoving disclination, and $\varphi_1$ is a small correction to
$\varphi_0$. The gradients of $\varphi_0$ are determined by Eq.
(\ref{a8}).

Linearizing the equations (\ref{veloc},\ref{phiphi}) in
$\varphi_1$, one obtains
\begin{eqnarray}
&& \eta\nabla^2 v_\alpha +\frac{K}{2}
\epsilon_{\alpha\beta}\nabla_\beta \nabla^2 \varphi_1
-K\nabla_\alpha\varphi_0 \nabla^2 \varphi_1
+\nabla_\alpha \left[\varsigma
- \frac{K}{2}(\nabla\varphi)^2 \right]= 0 \,,
\label{rg2} \\ &&
\nabla^2\varphi_1
-\frac{\gamma }{K}v_\alpha \nabla_\alpha\varphi_0
+\frac{\gamma}{2K}\epsilon_{\alpha\beta} \nabla_\alpha v_\beta
=- \frac{\gamma }{K}V_\alpha \nabla_\alpha\varphi_0 \,.
\label{rg3} \end{eqnarray}
Introducing a new variable $\chi=(K/\eta)\nabla^2\varphi_1$ we
rewrite Eqs. (\ref{rg2},\ref{rg3}) as
\begin{eqnarray} &&
\nabla^2 v_\alpha +\frac{1}{2}
\epsilon_{\alpha\beta}\nabla_\beta \chi
-\nabla_\alpha\varphi_0 \chi +\nabla_\alpha\varpi= 0 \,,
\label{rg4} \\ &&
\chi-\varGamma v_\alpha
\nabla_\alpha\varphi_0 +\frac{\varGamma}{2}
\epsilon_{\alpha\beta}\nabla_\alpha v_\beta
=-\varGamma V_\alpha\partial_\alpha\varphi_0 \,,
\label{rg5} \end{eqnarray}
where as above $\varGamma=\gamma/\eta$ and
$\varpi=\eta^{-1}\left[\varsigma-{K}/{2}(\nabla\varphi)^2\right]$.
It follows from Eq. (\ref{rg4}) and $\nabla_\alpha v_\alpha=0$ that
$\nabla^2\varpi=\nabla_\alpha\varphi_0\nabla_\alpha\chi$.
A solution of the system (\ref{rg4}-\ref{rg5}) can be written as
\begin{eqnarray}
v_\alpha = V_{\alpha }
+\epsilon_{\alpha\beta}\nabla_\beta\Omega \,,
\label{veom} \end{eqnarray}
where $V_\alpha$ is an obvious (due to Galilean invariance) forced
solution and the stream function $\Omega$ describes a zero mode of
the system (\ref{rg4}-\ref{rg5}). The system is homogeneous in $r$,
and therefore $\Omega$ is a sum of contributions which are
power-like functions of $r$.

Taking curl of Eq. (\ref{rg4}) we obtain
\begin{eqnarray}
-\nabla^4 \Omega -\frac{1}{2} \nabla^2 \chi
-\epsilon_{\gamma\alpha} \nabla_\alpha \varphi_0
\nabla_\gamma\chi=0 \,.
\label{rg15} \end{eqnarray}
Substituting $\chi$ expressed in terms of $\bm v$ from Eq.
(\ref{rg5}) into Eq. (\ref{rg15}), and using explicit expressions
for derivatives of $\varphi_0$, in the polar coordinates $(r,\phi)$
we obtain
\begin{eqnarray}
\left(1+\frac{\varGamma}{4} \right)\nabla^4 \Omega + s{\varGamma}
\left(\frac{2}{r^2}\partial^2_r \Omega -\frac{1}{r^2}\nabla^2 \Omega
-s\frac{1}{r^2}\partial^2_r \Omega
+s\frac{1}{r^3}\partial_r \Omega \right)=0\,.
\label{rg17} \end{eqnarray}
Solutions of Eq. (\ref{rg17}) are superpositions of terms $\propto
r^{\alpha+1}\exp(im\phi)$. Substituting this $r,\phi$-dependence
into Eq. (\ref{rg17}), one obtains the equation for $\alpha$, which
has the following roots
\begin{eqnarray} &&
\alpha=\pm\frac{1}{\sqrt 2} \left[2+2m^2
-s(1-s){\tilde\varGamma} \pm
\sqrt{\left(2+2m^2-s(1-s){\tilde\varGamma}\right)^2
-4 s{\tilde\varGamma}(m^2-1+s) -4(m^2-1)^2} \right]^{1/2} \,,
\label{rg19} \end{eqnarray}
where $\tilde\varGamma=\varGamma(1+{\varGamma}/{4})^{-1}$. Hence
$0<\tilde\varGamma<4$ for any $\gamma$ and $\eta$. Evidently, all
the roots (\ref{rg19}) are real. Let us emphasize that there is no
solution $\alpha=0$ (corresponding to a logarithmic in $r$ behavior
of the velocity) among the set (\ref{rg19}). The first angular
harmonic with $|m|=1$ is of particular interest because far from
the disclination $\varphi_1=ur\sin\phi$ and $\Omega=-Vr\sin\phi$.
If $\varGamma$ is small, there is a pair of small exponents
\begin{eqnarray}
\alpha=\pm\alpha_1\,, \qquad
\alpha_1=s \sqrt{\varGamma}\,/2 \,,
\label{rg20} \end{eqnarray}
for $m=\pm1$. Otherwise, for any other relevant $m$ (terms with
$m=0$ are forbidden because of the symmetry), exponents
(\ref{rg19}) have no special smallness.

We established that $\Omega$ is a superposition of the terms
$\propto r^{\alpha+1}\exp(im\phi)$ with $\alpha$ determined by Eq.
(\ref{rg19}). Then the velocity can be found from Eq. (\ref{veom}).
To avoid a singularity in the velocity at small $r$, one should
keep contributions with positive $\alpha$ only. By other words, one
can say, that the velocity field contains contributions with all
powers $\alpha$ (\ref{rg19}) but the factors at the terms with
negative $\alpha$ are formed at $r\sim a$ (where $a$ is the
disclination core radius), and therefore the corresponding
contributions to the velocity are negligible at $r\gg a$ (this
statement should be clarified and refined for small negative
exponents $-\alpha_1$ in the limit of small $\varGamma$, see below,
Subsection \ref{sub:extr}). We conclude, that the correction to
$\bm V$ in the flow velocity $\bm v$, related to $\Omega$ in Eq.
(\ref{veom}), is negligible at $r\sim a$. Thus we come to the
non-slipping condition for the disclination motion: the
disclination velocity $\bm V$ coincides with the flow velocity $\bm
v$ at the disclination position.

Next, to find $\varphi$, one should solve the equation
$(K/\eta)\nabla^2\varphi=\chi$, where $\chi$ is determined from Eq.
(\ref{rg5}). Then, besides the part determined by the velocity,
zero modes of the Laplacian could enter $\varphi_1$. The most
dangerous zero mode is $Uy$, since it produces a non-zero momentum
flux (and associated to it the Magnus force)
\begin{eqnarray}
\oint\mbox d r_\alpha\,\epsilon_{\alpha\beta}
T_{\beta\gamma}\sim KU \,,
\nonumber \end{eqnarray}
to the disclination core. However, because of the condition
$\alpha\neq0$, all the contributions to the velocity correspond to
zero viscous momentum flux to the origin. Consequently, it is
impossible to compensate the Magnus force by other terms. The above
reasoning leads us to the conclusion that the factor $U$ (and
therefore the Magnus force) must be zero. Thus, $\varphi_1$
contains only terms, proportional to $r^{\alpha+1}$, with
$\alpha>0$. This conclusion is related to the fact that for free
standing liquid crystalline films any distortion of the bond angle
unavoidably produces hydrodynamic back-flow motions (i.e. ${\bm
v}\neq0$). Unlike free standing films, for the liquid crystalline
films on substrates (Langmuir films) hydrodynamic motions
(back-flows) are strongly suppressed by the substrate, and a
situation when the back-flow is irrelevant for the disclination
motion could be realized.

\subsection{Remote Region}
\label{sub:far}

Let us consider the region $r\gg u^{-1}$, where we can write
$\varphi=uy+\tilde\varphi$ and linearize the system of equations
(\ref{veloc},\ref{phiphi}) with respect to $\tilde\varphi$. Then we
get the system of linear equations for $\bm v$ and $\tilde\varphi$:
\begin{eqnarray} &&
\nabla^2 v_\alpha+\frac{K}{2\eta}
\left(\epsilon_{\alpha\beta}\nabla_\beta
\nabla^2 \tilde\varphi
-2u_\alpha\nabla^2 \tilde\varphi\right)
+\nabla_\alpha\varpi=0 \,,
\nonumber \\ &&
(\nabla^2 + 2 p \partial _x) \tilde\varphi +
\frac{\gamma}{2K} (\epsilon_{\alpha\beta}
\nabla_\alpha v_\beta-2u v_y )=0\,,
\label{grr1} \end{eqnarray}
where $p=V\gamma/2K$. Taking curl of the first equation and
excluding the Laplacian, we obtain
\begin{eqnarray}
\epsilon_{\beta\alpha} \nabla_\beta v_\alpha=
\frac{K}{2\eta}\left[
\left( \nabla^2+2u\partial_x\right) \tilde\varphi
+\Phi\right] \,,
\label{ggr1} \end{eqnarray}
where $\Phi$ is a harmonic function. In terms of the harmonic
function $\Phi$ the system (\ref{grr1}) is reduced to
\begin{eqnarray}
\left[(1+\varGamma/4)\nabla^4+2p\nabla^2\partial_x
-\varGamma u^2\partial_x^2\right]\tilde\varphi=
(\varGamma/2) u \partial_x \Phi \,.
\label{gr1} \end{eqnarray}
The equation (\ref{gr1}) can be written as
\begin{eqnarray} &&
(\nabla^2+2k_1\partial_x)
(\nabla^2-2k_2\partial_x)\tilde\varphi=
(\tilde\varGamma/2)u\partial_x \Phi \,,
\label{far1} \\ &&
k_{1,2}=\frac{1}{2(1+\varGamma/4)}\left(
\sqrt{p^2+\varGamma(1+\varGamma/4) u^2}\,\pm p\right)\,.
\label{far2} \end{eqnarray}
The quantities $k_1$ and $k_2$ have the meaning of characteristic
wave vectors. We conclude from Eq. (\ref{far1}) that zero modes of
the operator in the left hand side of the equation are proportional to
\begin{eqnarray} &&
\exp(-k_1r-k_1x)\,, \qquad \exp(-k_2r+k_2x) \,,
\nonumber \end{eqnarray}
that is they are exponentially small everywhere, except for narrow
angular regions near the $X$-axis. The behavior of the zero modes
inside the regions is power-like in $r$. Besides, there is a
contribution to $\tilde\varphi$ related to the harmonic function
$\Phi$. It contains a part which decays as a power of $r$ (the
leading term is $\propto r^{-1}$) at large $r\gg u^{-1}$. This
solution is examined in more detail in Appendix \ref{ap:expli}.

\section{Different regimes governed by $\varGamma$}
\label{cases}

A character of the velocity and the bond fields around the moving
disclination is sensitive to the ratio of the rotational and the
shear viscosity coefficients $\varGamma=\gamma/\eta$. In this
section we examine different cases, depending on the $\varGamma$
value.

\subsection{The case $\varGamma\gtrsim 1$}
\label{sub:normal}

We start to treat different regimes of mobility with the most
likely case $\varGamma\gtrsim 1$. If $\varGamma\sim1$ then the
factor $C$ in Eq. (\ref{final}) is of order unity and $u\sim p$.
Then we find from Eqs. (\ref{far2}), that $k_1,k_2\sim u$. It is a
manifestation of the fact, that there is the unique characteristic
scale in this case, which is $u^{-1}$. Then one can estimate
$\tilde\varphi$ by matching at $r\sim u^{-1}$ of the solutions in
the regions near the disclination and far from it. We conclude,
that it is a function of a dimensionless parameter $ur$; the
function is of the order of unity, when its argument $ur$ is of the
order of unity.

For large $\varGamma$ there remains the unique characteristic
scale $u^{-1}$ and, consequently, in this case still $C\sim1$. To
prove the statement, we treat first small distances $r\ll u^{-1}$.
As we have shown in Subsection \ref{sub:near}, the corrections
$\varphi_1$ and $\delta{\bm v}$, to $\varphi_0$ and $\bm V$
respectively, are expanded into the series over the zero modes
characterized by the exponents (\ref{rg19}). Particularly, for
$m=1$ we can write $\varphi_1\sim uy (ur)^\alpha$. In the large
$\varGamma$ limit the exponents $\alpha $ (\ref{rg19}) are regular
since $\tilde\varGamma\to4$. From (\ref{rg19}) we have
$\alpha_1\sim 1$, and for this case
\begin{eqnarray}
\chi\sim \frac{K}{\eta r^2}uy (ur)^{\alpha_1} \,.
\nonumber \end{eqnarray}
Comparing equations (\ref{rg4}) and (\ref{rg5}), we conclude that
at large $\varGamma$ the term with $\chi$ in the second equation
can be omitted and therefore the equation is a constraint imposed
on the velocity. Then the first equation gives
\begin{eqnarray}
|\delta v|\sim \frac{K}{\eta r}uy (ur)^{\alpha_1}  \,.
\nonumber \end{eqnarray}
The disclination velocity can be found now from the relation
$V\sim|\delta v|$ at the scale $u^{-1}$, that is $p\sim\varGamma
u$, or $C\sim 1$. The complete analysis covers also the remote
region. At the condition $p\sim \varGamma u$ both $k_{1,2}\sim
u^{-1}$. Then using the procedure of Appendix \ref{ap:expli}, one
can prove that the solutions from two regions can be matched at
$r\sim u^{-1}$, and thus there are no new characteristic scales,
indeed. We also note that the rotational viscosity $\gamma$ is
excluded from the hydrodynamic equations at large $\varGamma$.
Although this is not true inside the disclination core provided
that one can employ hydrodynamics there (see Appendix
\ref{ap:core}), boundary conditions for $\bm v$ and $\varphi$ on
the core reveal no dramatic behavior. Consequently, it is the shear
viscosity alone that determines the disclination mobility, which
means that $C\sim 1$.

Thus we can say that the limit $\varGamma\to\infty$ is trivial: no
any additional feature appears in comparison with $\varGamma\sim1$.
However, it is not the case for small $\varGamma$, since for
$\varGamma\ll1$ it turns out that $u\gg p$. We study this case in
the next Subsection.

\subsection{Small $\varGamma$}
\label{sub:small}

Here, we consider the case $\varGamma\ll1$. This limit is attained
at anomalously large $\eta$, so that $K\rho/\eta^2$ should be still
treated as the smallest dimensionless parameter. That justifies use
of the same equations (\ref{veloc},\ref{phiphi}), as in the
previous subsections.

At $r\ll u^{-1}$ the analysis of Subsection \ref{sub:near} is
correct. As we noted, contributions to $\bm v$ and $\varphi_1$,
related to the modes with negative $\alpha$, should not be taken
into account there. At $\varGamma\ll1$ the leading role is played
by the mode with the smallest exponent
$(\alpha_1=s\sqrt\varGamma/2)$, because the presence of modes with
positive exponents $\alpha\sim 1$ would contradict the condition of
smooth matching at $r\sim u^{-1}$. Strictly speaking, neglecting a
small negative exponent $-\alpha_1$ is correct at the condition
$\alpha_1|\ln(ua)|\gg1$, where $a$ is the core radius of the
disclination. In this subsection we consider just this case. The
opposite case, that we will term it as extremely small $\varGamma $
limit, is analyzed in Subsection \ref{sub:extr}. At $r\ll u^{-1}$
we can write
\begin{eqnarray}
\varphi_1\sim uy (ur)^{\alpha_1} \,, \qquad
V-v_x \sim \alpha_1 u \frac{K}{\gamma}
(ur)^{\alpha_1} \,,
\label{sma1} \end{eqnarray}
where the coefficient at $y(ur)^{\alpha_1}$ is determined from
matching at $r\sim u^{-1}$ where $\nabla\varphi\sim 1/r$. Analogous
matching $V-v_x\sim V$ at $r\sim u^{-1}$ gives $V\sim \alpha_1
uK/\gamma$. The relation can be rewritten as $p\sim \alpha_1 u\ll
u$. Therefore we conclude $C\sim 1/\sqrt\varGamma$.

The relation $p\sim\sqrt\varGamma\,u$ leads to $k_{1,2}\sim p\ll u$, as
follows from Eq. (\ref{far2}). By other words, a new scale $p^{-1}$
(different from $u^{-1}$) appears in the problem. Therefore a detailed
investigation of the far region $r\gg u^{-1}$ is needed, to establish
an $\bm r$-dependence of the bond angle $\varphi$ and the velocity
field $\bm v$ there. This investigation can be based on the equations,
formulated in Subsection \ref{sub:far}, which are correct irrespective
to the value of $pr$.

Explicit expressions, describing the velocity and the angle, are
presented in Appendix \ref{ap:expli}. They contain three
dimensionless functions $\zeta_1(\nabla/u)$, $c_1(\nabla/u)$ and
$c_2(\nabla/u)$. Really, at $ur\gg1$ one can keep only zero terms
of the expansions of these functions into the Taylor series. Among
these three coefficients only one is independent, see Eq.
(\ref{flux3}). Therefore the general solution could be expressed
with a single free parameter, $\zeta\equiv
\zeta_1(0)$. The procedure corresponds to the following construction of
solutions of the equations of motion (\ref{grr1}) in the region
$ur\gg1$. We have to match the solutions in the outer (far from the
disclination) and in the inner (close to the disclination) regions
at $ur\simeq1$. Technically the matching is equivalent to the
appropriate boundary conditions for the outer problem at
$ur\simeq1$, and these boundary conditions can be replaced formally
by the local source terms in the equations, acting at $ur\simeq1$.
One can expand these sources in the standard multipolar series.
Thus we come to the expansion over the gradients of
$\delta$-function. The gradients scale as $u$, and therefore
$\zeta$, $c_1$, $c_2$ are dimensionless functions of the
dimensionless ratio $\nabla/u$.

To establish asymptotic behavior of the angle $\varphi$ and of the
velocity $\bm v$, let us first consider the region $u^{-1}\ll r\ll
p^{-1}$. Then we derive from Eqs. (\ref{aaa2},\ref{BC},\ref{flux3})
\begin{eqnarray} &&
v_x=\frac{K(2s-\zeta)}{\gamma u}
k_1k_2\ln(pr) \,.
\label{sma3} \end{eqnarray}
Here we presented only the leading logarithmic contribution of
the zero harmonic in $v_x$. Matching the velocity
derivatives, determined by Eqs. (\ref{sma1},\ref{sma3}), at $r\sim
u^{-1}$, we find $\zeta\sim1$ (we imply $s\sim 1$).
Using the expressions
(\ref{aaa1},\ref{BC},\ref{flux3}), we obtain for the region
$u^{-1}\ll r\ll p^{-1}$
\begin{eqnarray} &&
\varphi=\varphi_0+uy+spy\ln(pr) \,.
\label{sma2} \end{eqnarray}
We see, that there is only a small correction to the simple expression
$\varphi_0+uy$ there, since $p\ll u$.

In the region $pr\gg1$, the expressions for the angle $\varphi$ and
the velocity $\bm v$ are more complicated. Using the formulas
(\ref{aaa1},\ref{aaa9},\ref{aaa2},\ref{BC}), one obtains
\begin{eqnarray} &&
\partial_x\varphi=-s\sqrt{\frac{\pi}{2}}
\left[c_1\sqrt{k_1}\,\exp(-k_1r-k_1x)
+c_2\sqrt{k_2}\,\exp(-k_2r+k_2x)\right]
\frac{y}{r^{3/2}}-\frac{\zeta}{2}\frac{y}{r^2} \,,
\label{sma4} \\ &&
\partial_y\varphi=u-2s\sqrt{\frac{\pi}{2}}
\left[c_1\sqrt{k_1}\,\exp(-k_1r-k_1x)
-c_2\sqrt{k_2}\,\exp(-k_2r+k_2x)\right]
\frac{1}{r^{1/2}}+\frac{\zeta}{2}\frac{x}{r^2} \,,
\label{sma5} \\ &&
v_y=\frac{K}{\gamma u}\left\{
2s\sqrt{\frac{\pi}{2}}\left[
c_1 k_2\sqrt{k_1}\,\exp(-k_1r-k_1x)
-c_2k_1\sqrt{k_2}\,\exp(-k_2r+k_2x)\right]\frac{y}{r^{3/2}}
-p\zeta \frac{y}{r^2}\right\} \,,
\label{sma6} \\ &&
v_x=-\frac{K}{\gamma u}\left\{\sqrt{\frac{\pi}{2}}
s\varGamma u^2\left[
\frac{c_1}{\sqrt{k_1r}}\,\exp(-k_1r-k_1x)
+\frac{c_2}{\sqrt{k_2r}}\,\exp(-k_2r+k_2x)\right]
+p\zeta \frac{x}{r^2}\right\} \,,
\label{sma8} \end{eqnarray}
where $c_1\sim 1$ and $c_2\sim 1$ are determined by Eq.
(\ref{flux3}) (we omitted the argument $0$ to simplify the
notations). The expressions (\ref{sma4},\ref{sma5},\ref{sma6},\ref{sma8})
contain two types of terms: isotropic and anisotropic ones. The
anisotropic contributions are essential only in the narrow angular
regions near the $X$-axis, where they dominate. It is worth to note
the very non-trivial structure of the flow, where the isotropic
flux to the origin is compensated due to the anisotropic terms.

The expressions found in this subsection generalize the famous Lamb
solution for the hydrodynamic flow around a hard cylinder, (see,
e.g., \cite{BA70,Lamb,LL8}) where far from the cylinder the
velocity field is exponentially small everywhere except in the wake
of the corps, i.e. in the very narrow angular sector (``tail").
Disclination motion in liquid crystalline films can be regarded as
motion of a cylinder framed by a ``soft" (i.e. deformable)
orientational field $\varphi$. Due to the additional (with respect
to the classical Lamb problem) degree of freedom, our solution has
two tails around the moving disclination: wake, beyond the
disclination, and precursor in front of it. In fact, both degrees
of freedom (the flow velocity and the bond angle) are relevant.

\subsection{Extremely small $\varGamma$}
\label{sub:extr}

At the above analysis we implied the condition
$\alpha_1|\ln(ua)|\gg1$ (remind that $\alpha_1=s \sqrt\varGamma/2$
at small $\varGamma$), imposing a restriction from below to
$\varGamma$ at a given $u$. If $\alpha_1|\ln(ua)|\ll1$, then the
terms with both $\alpha=\pm\alpha_1$, determined by Eq.
(\ref{rg20}), should be taken into account near the disclination,
that leads to a logarithmic behavior of the correction $\varphi_1$
to $\varphi_0$ there:
\begin{eqnarray} &&
\varphi_1\sim uy \ln(r/a) |\ln (au)|^{-1} \,,
\label{sma7} \end{eqnarray}
instead of Eq. (\ref{sma1}). Matching the derivatives of
expressions (\ref{sma2},\ref{sma7}) at $r\sim u^{-1}$ gives $p\sim
u|\ln(au)|^{-1}$. By other words, $C\sim [\varGamma \ln(au)]^{-1}$.
This case corresponds formally to the limit $\eta\to\infty$ in our
equations, when one can drop the back-flow hydrodynamic velocity in
the equation for the bond angle. The situation was examined in the
works \cite{BP88,PR90,RK91,DE96}. We present a simple analysis of
the case in Appendix \ref{ap:logre}. Note also that, really, there
is no crossover at $r\sim u^{-1}$ in the bond angle behavior in
this situation.

Let us clarify the question, concerning the Magnus force in this case.
In accordance with Eq. (\ref{sma7}) the reactive momentum flux to the
disclination core is
\begin{eqnarray}
\oint\mbox d r_\alpha\,\epsilon_{\alpha\beta}
T_{\beta\gamma} \sim Ku
\ln(r/a) |\ln (au)|^{-1} \,.
\nonumber \end{eqnarray}
Thus, the flux is $r$-dependent, tending to zero at $r\to a$. This
reactive momentum flux is compensated by the viscous momentum flux,
(related to derivatives the flow velocity $\bm v$), which is
non-zero in this case because of the logarithmic in $r$ behaviour
of the flow velocity near the disclination. The flow velocity can
be found from Eqs. (\ref{rg2},\ref{sma7}), it behaves as
\begin{eqnarray}
v_\alpha\sim \frac{Ku}{\eta|\ln(au)|}\,
\epsilon_{\alpha\beta}\nabla_\beta[y\ln^2(r/a)] \,,
\nonumber \end{eqnarray}
what is a generalization of the Stokes-Lamb solution \cite{BA70,Lamb}.
However unlike the Lamb problem (a hard cylinder moving in a
viscous liquid), in our case $|\bm V-\bm v(r=a)|\sim V$, i.e. we
have a slipping on the core of the moving disclination. This
slipping seems natural in the limit of extremely small values of
$\varGamma$, corresponding to the limit $\eta\to\infty$, that is to
a strongly suppressed hydrodynamic flow. Physically, the property
means that the disclination can not be understood as a hard
impenetrable object in the case. It is worth noting also that found
above logarithmic behavior looks similar as a general feature of
the two-dimensional hydrodynamic motion, which comes from the well
known fact (see e.g., \cite{BA70,Lamb,LL8}) that for
two-dimensional laminar flow even for a small Reynolds number one
cannot neglect nonlinear terms, which become relevant for large
enough distances. However, in our case these non-linear terms do
not come from the convective hydrodynamic nonlinearity, they come
from the nonlinear over $\varphi$ terms in the stress tensor
(\ref{lli1}).

An explicit expression for $\varphi$ and its asymptotics,
corresponding to the considered case, are found in Appendix
\ref{ap:logre}. An expression for the flow velocity field, excited
by the disclination motion, is derived in Appendix \ref{ap:extreme}.

\section{Conclusion}
\label{discussion}

Let us sum up the results of our paper. To understand physics,
underlying the freely suspended film dynamics, we studied the
ground case, namely, a single disclination motion in a thin hexatic
(or smectic-$C$) film, driven by a inhomogeneity in the bond (or
director) angle. We investigated the uniform motion (i.e. one with
a constant velocity). For this case we derived and solved equations
of motion and found bond angle and hydrodynamic velocity
distributions around the disclination. It allows us to relate the
velocity of the disclination $V$ to the bond angle gradient
$u=|\nabla\varphi|$ in the region far from the disclination. It is
in fact the reason, why so much efforts are needed: the full set of
the equations should be solved everywhere, not locally only. We
established the proportionality coefficient $C$ (see Eq.
(\ref{final})) in this non-local relationship which has a meaning
of the effective mobility coefficient. The coefficient $C$ depends
on the dimensionless ratio $\varGamma$ of rotational $\gamma$ and
shear viscosity $\eta$ coefficients.

There is little experimental knowledge as far as the values of
the coefficients $\gamma$ and $\eta$ in hexatic and smectic-$C$
films are concerned. It is believed generally that the
corresponding values in a film (normalized by its thickness) and in
a bulk material are not very different \cite{BC96,OP00}, and in
this case we are in the regime of $\varGamma\simeq 1$, when the
coefficient $C$ is of the order of unity. However $\varGamma\ll1$
case is not excluded from the both theoretical and material science
points of view. We found for small $\varGamma\ll1$ case the
coefficient $C\sim1/\sqrt\varGamma$. We established a highly
non-trivial behaviour of the flow velocity and of the bond angle,
which is power-like in $r$ near the disclination and extremely
anisotropic far from it. And only for the case of extremely small
$\varGamma\ll1/\ln^2(ua)$ (where $a$ is the disclination core
radius), we found a logarithmic behavior $C\sim
[\varGamma\ln(ua)]^{-1}$. The main message of our study is that the
hydrodynamic motion (that is the back-flow), unavoidably
accompanying any defect motion in liquid crystals, plays a
significant role in the disclination mobility. Experimental
evidence (see, e.g., the recent publication \cite{AT00}) shows that
it is indeed the case.

Our consideration can be applied to the motion of a disclination
pair with opposite signs. In this case the role by the scale
$u^{-1}$ is played by the distance $R$ between the disclinations.
Then we find in accordance with Eq. (\ref{final}) that $\partial_t
R\propto R^{-1}$ without a logarithm (provided the twist viscosity
coefficient $\gamma$ is not anomalously small, see Subsection
\ref{sub:extr} for the quantitative criterion). This conclusion is
confirmed by results of numerical simulations for $2d$ nematics
\cite{FU98,DO00,DO01,TD02}. The authors of the works consider the
equations of motion in terms of the tensor order parameter,
consistently taking into account the coupling between the
disclination motion and the hydrodynamic flow. They computed
dynamics of the disclination pair annihilation and found that the
distance $R$ between the disclinations scales with time $t$ as
$t^{1/2}$, without logarithmic corrections (as it follows from our
theoretical consideration for any (excepting extremely small) value
of the parameter $\varGamma $. Unfortunately we did not find in the
publications \cite{DO00,DO01,TD02} the magnitudes of the the shear
viscosity which was used in the simulations. Lacking sufficient
data on the values of $\gamma$ and $\eta$ we can at present discuss
only the general features of the disclination dynamics. For
instance the authors \cite{TD02} found numerically in one-constant
approximation the asymmetry of the disclination dynamics with
respect to the sign of the topological charge ($s=\pm 1/2$). In our
approach the asymmetry naturally appears from nonlinear terms in
the stress tensor (\ref{a7}) and from the first term in the
right-hand side of Eq. (\ref{IIi2}) responsible for the different
coupling of orientational and hydrodynamic flow patterns for
positive and negative disclinations. This results in the fact that
smaller positive exponents in Eq. (\ref{rg19}) (corresponding to
the minus in the brackets) for each $m$ are larger if $s=1/2$ than
those for $s=-1/2$. Thus, the disclination with $s=1/2$ exerts a
stronger influence on the flow velocity, the qualitative conclusion
that was arrived in \cite{TD02}.

Although the theory, presented in the paper, is valid for free
standing hexatic or smectic-$C$ liquid crystalline films, the
general scheme can be applied to Langmuir films, that is for the
liquid crystalline films on solid or liquid substrates. Since the
Langmuir film is arranged on the substrate surface, then any its
hydrodynamic motion is accompanied by the substrate motion.
Therefore for solid substrates a situation, when the hydrodynamic
back-flow is irrelevant for the disclination dynamics, could be
realistic. In Subsection \ref{sub:extr} (see also Appendix
\ref{ap:logre}) we examine this limit, and reproduce the results of
the works \cite{RK91,BP88,PR90,DE96}, where the hydrodynamic
back-flow was neglected from the very beginning. As it concerns to
the Langmuir films on a liquid substrate, this case requires a
special investigation, however the approach and main ideas of our
paper could be useful as well.

Our results can be tested directly by comparing with experimental
data for smectic-$C$ films. The hexatic order parameter, which has
sixfold local symmetry is not coupled to the light in any simple
way (and therefore ideal hexatic disclinations are hardly observed
in optics). However it is possible to observe the core splitting of
the disclinations in tilted hexatic smectic films \cite{DP86}.
Indeed, due to discontinuity the tilt direction (which is locked to
the bond direction) one can observe hexatic order and hexatic
disclinations indirectly. The second possibility of checking our
theoretical results is the classical light-scattering (where in
typical experiments wave vectors are $q=10^2\div10^4cm^{-1}$, and
the frequency is $\omega\lesssim10^8s^{-1}$). For a reasonably
thick film the power spectrum of light scattering can have some
additional structure that reveals the disclination properties (e.g.
defects are thought to be relevant to the very low-frequency noise
observed in thin films). Such kind of experimental studies are
highly desirable.

\acknowledgments

The research described in this publication was made possible in
part by RFFR Grant 00-02-17785 and INTAS Grant 30-234. SVM thanks
the support of this work by the Deutsche Forschungsgemeinschaft,
Grant KO 1391/4. Fruitful discussions with V. E. Zakharov, E. A.
Kuznetzov, G. E. Volovik, and N. B. Kopnin are gratefully
acknowledged. We thank as well E. B. Sonin for sending us reprint
of his paper \cite{SO97}.

\appendix

\section{}

Here we present some technical details, concerning the velocity and
the bond angle fields around the disclination. The case of small
$\varGamma$ is mostly treated, which is especially rich from the point of
view of the space structure of the fields.

\subsection{Distances far from the disclination}
\label{ap:expli}

Here we derive some results for the far from the disclination
region which are afterwards applied to the case of small $\varGamma$
considered in Subsection \ref{sub:small}.

Let us examine the harmonic function $\Phi$ in Eqs. (\ref{ggr1}).
Since the function is analytic in the region $r>u^{-1}$, it can be
expanded over derivatives of $\ln r$ there. Next, due to the
symmetry of the problem, $\Phi$ should be antisymmetric in $y$.
Therefore at least one derivative $\partial_y$ should be present in
each term of the expansion, that is
\begin{eqnarray}
\Phi= u \hat\zeta_1\partial_y \ln r \,,
\label{far4} \end{eqnarray}
where $\hat\zeta_1=\zeta_1(\nabla/u)$, and $\zeta_1(z)$ is a series of
$\bm z$, converging in a circle with radius of the order of unity.
The expansion coefficients in the series $\zeta_1(\nabla/u)$ are
determined by matching at $r\sim u^{-1}$ with the inner problem.

Because of the symmetry, the angle $\tilde\varphi$ can be
represented in the following form:
\begin{eqnarray} &&
\partial_x\tilde\varphi=
\partial_y B \,, \qquad
\partial_y\tilde\varphi=
-(H+\partial_x B)\,, \qquad
\nabla^2 B+\partial_x H=0\,.
\label{aaa1} \end{eqnarray}
The latter equation is the condition
$\epsilon_{\alpha\beta}\nabla_\alpha\nabla_\beta\tilde\varphi=0$.
Note that $\nabla^2\tilde\varphi=-\partial_yH$. In the far from the
disclination region we can use Eqs. (\ref{grr1},\ref{ggr1}). The
incompressibility condition $\nabla_\alpha v_\alpha=0$ must be
taken into account as well. Then one obtains expressions for the
velocity in terms of $B$ and $H$:
\begin{eqnarray} &&
\mathrm{curl}\ \bm v =\frac{K}{2\eta}\partial_y
\left[-H+2uB+u\hat \zeta_1 \ln(pr) \right]\,,
\nonumber \\ &&
v_y =\frac{K}{\gamma u}\partial_y
\left\{-H+2pB+\frac{\varGamma}{4}
\left[-H+2uB+u\hat \zeta_1 \ln (pr)\right] \right\}\,,
\label{aaa9} \\ &&
v_x =\frac{K}{\gamma u}
\left\{\partial_x\left\{-H+2pB
+\frac{\varGamma}{4}\left[-H+2uB+u\hat \zeta_1
\ln (pr)\right]\right\}
-\frac{\varGamma u}{2}\left[-H+2uB+u\hat \zeta_1
\ln (pr) \right]\right\}\,,
\label{aaa2} \end{eqnarray}
Solutions of Eq. (\ref{gr1}) imply:
\begin{eqnarray} &&
B=s\biggl[
\hat c_1 K_0(k_1r)e^{-k_1x} + \hat c_2 K_0(k_2r)e^{k_2x}
\biggr] -\frac{1}{2} \hat\zeta_1\ln(pr)\,,
\nonumber \\ &&
H=2s\left[
k_1\hat  c_1 K_0(k_1r)e^{-k_1x}
- k_2\hat c_2 K_0(k_2r)e^{k_2x}\right] \,.
\label{BC} \end{eqnarray}
Here the particular representation (\ref{far4}) is used and an
arbitrary function of $y$ which can contribute to $H$ is chosen to be
zero because $\nabla\tilde\varphi\to 0$ (and hence $H\to 0$) as
$r\to\infty$. Above $\hat c_1$ and $\hat c_2$ are dimensionless
differential operators which can be represented as Taylor series over
$\nabla/u$, i.e. $c_1(\nabla/u)$ and $c_2(\nabla/u)$. These functions
must scale with $u$ since the functions have to be found
from the matching at $r \simeq u^{-1}$.

Additionally, there are two conditions on the variables in the
region $ur\gg1$. First, the correct circulation around the origin
leads to the effective $\delta$-functional term in Eq. (\ref{aaa1}):
\begin{eqnarray} &&
\nabla^2 B+\partial_x H
=-2 {\pi}s\delta(\bm r)\,.
\label{circ} \end{eqnarray}
The second condition is the absence of the flux to the origin:
\begin{eqnarray} &&
\int \mbox d\phi\, v_r(r,\phi)=0\,.
\label{flux} \end{eqnarray}
The relations (\ref{circ},\ref{flux}) lead to the conditions
\begin{eqnarray} &&
c_1(0)+c_2(0) + \zeta_1(0)/2s=1\,,
\label{circ1} \\ &&
\left(1+\frac{\varGamma}{4}\right)
\left[k_1c_1(0)-k_2c_2(0)\right]
-\left(p+\frac{\varGamma u}{4}\right)\left[c_1(0)
+c_2(0) + \zeta_1(0)/2s\right]
+ \frac{\varGamma u}{8s}\zeta_1(0)=0\,.
\label{flux1} \end{eqnarray}
At small $\varGamma$ the solution of the equations
(\ref{circ1},\ref{flux1}) is
\begin{eqnarray} &&
\zeta_1(0)=\zeta\,, \quad
c_1(0)=\frac{k_1-\zeta k_2/2s}{k_1+k_2}\,, \quad
c_2(0)=\frac{k_2-\zeta k_1/2s}{k_1+k_2} \,.
\label{flux3}\end{eqnarray}
We also assumed that $\zeta\lesssim 1$, which is justified in
Subsection \ref{sub:small}.

\subsection{Suppressed Flow}
\label{ap:logre}

Here we demonstrate, for convenience of references, how is it
possible to find the disclination velocity $V$ provided the
hydrodynamic velocity ${\bm v}$ is negligible (say, due to a
substrate friction). We reproduce the results of the works
\cite{BP88,PR90,RK91,DE96}.

In the absence of the hydrodynamic flow the equation for the angle
$\varphi$ is purely diffusive:
\begin{eqnarray} &&
\gamma \partial_t\varphi
=K\nabla^2\varphi \,,
\label{dif1} \end{eqnarray}
as follows from Eq. (\ref{IIi2}) at $\bm v=0$. We assume that
$\varphi\to uy$ if $r\to\infty$. The disclination motion is forced
by the ``external field'' $u$. Next, we are looking for a solution
$\varphi(t,x,y)=\varphi(x-Vt,y)$. Then we get from Eq. (\ref{dif1})
\begin{eqnarray} &&
2p\partial_x\varphi+\nabla^2\varphi=0 \,,
\quad {\rm where}\quad  2p=\gamma  V/K \,.
\label{dif2} \end{eqnarray}
Below we consider a solution corresponding to a single disclination
with the circulation
\begin{eqnarray} &&
\oint \mbox d\bm r\, \nabla \varphi=2\pi s\,,
\label{dif25} \end{eqnarray}
where the integral is taken along a contour, surrounding the
disclination in the anticlockwise direction. The quantity $s$ in
Eq. (\ref{dif25}) is an arbitrary parameter (which is equal to $\pm
1/6$ for hexatics and $\pm 1/2$ for nematics). For a suitable
solution of Eq. (\ref{dif2}), corresponding to Eq. (\ref{dif25}),
we have
\begin{eqnarray} &&
\partial_x\varphi
=\partial_y\int\frac{{\rm d}^2q}{2\pi}
\frac{1}{q^2-2ipq_x}\exp(i{\bm q}{\bm r})
=s\exp(-px)\partial_yK_0(pr) \,.
\label{dif3} \end{eqnarray}
This derivative tends to zero at $r\to\infty$ as it should be.

The expression (\ref{dif3}) does not determine $\varphi$ itself
unambiguously, since $\partial_x(uy)=0$ and therefore one can add
an arbitrary function $uy$ to a solution and it gives a new
solution then. Note that $uy$ is a zero mode of the equation
(\ref{dif2}). Thus the solution can be written as
\begin{eqnarray} &&
\varphi=\phi+uy \,, \qquad
\phi(x',y)=-s\int_{x'}^\infty{\rm d}x\,
\exp(-px)\partial_yK_0(pr) \,,
\label{diff} \end{eqnarray}
where $\phi$ tends to zero if $r\to\infty$. To relate $u$ in Eq.
(\ref{diff}) and $p$, one must know boundary conditions at $r\to0$,
really, at $r\sim a$, where $a$ is the core radius. At small $r$
$\varphi$ can be written as a series
$\varphi=\varphi_0+\varphi_1+\dots$, where $\varphi_0$ corresponds
to an motionless disclination, $\varphi_1$ is the first correction
to $\varphi_0$, related to its motion. Matching with the inner
problem gives
\begin{eqnarray} &&
\nabla\varphi_1(a)\sim p \,,
\label{dif9} \end{eqnarray}
since the solution for the order parameter inside the core is an
analytical function of $r/a$, and the expansion over $p$ is a
regular expansion over $pa$.

Expanding Eq. (\ref{dif3}) over $p$ we get at $pr\ll1$
\begin{eqnarray} &&
\frac{1}{s}\partial_x\varphi\approx
-\frac{y}{r^2}+\frac{pxy}{r^2} \,.
\nonumber \end{eqnarray}
Therefore in accordance with Eq. (\ref{diff}) one gets
with the logarithmic accuracy
\begin{eqnarray} &&
\varphi_1=spy\ln(pr)+uy \,.
\label{dif11} \end{eqnarray}
Using now the boundary condition (\ref{dif9}) one gets with the same
logarithmic accuracy
\begin{eqnarray} &&
u=sp\ln\left(\frac{1}{pa}\right) \,.
\label{dif12} \end{eqnarray}
It can be rewritten as
\begin{eqnarray} &&
V=\frac{2Ku}{s\gamma \ln(1/pa)} \,.
\label{dif16} \end{eqnarray}

The same answer (\ref{dif16}) can be found from the energy dissipation
balance. First of all, we can find the energy $E$ corresponding to the
solution (\ref{diff}):
\begin{eqnarray} &&
E=\int {\rm d}^2 r\, \frac{K}{2}
(\nabla\varphi)^2=K\int {\rm d}^2 r\,
\left[ \frac{1}{2}u^2
+\frac{1}{2}(\nabla\phi)^2
+u\partial_y\phi \right] \,.
\label{dif51} \end{eqnarray}
Here the first term is the energy of the external field and the second
term is the energy of the disclination itself. Obviously, both the
terms are independent of time, and only the last cross term depends on
the time. For $|x-Vt|\gg p^{-1}$
\begin{eqnarray} &&
\int_{-\infty}^\infty{\rm d}y\,
\partial_y\phi=\left\{
\begin{array}{ccc}
0 & {\rm if} & x>V t \,, \\
-2\pi s & {\rm if} & x<Vt \,.
\end{array} \right.
\nonumber \end{eqnarray}
Therefore we obtain from Eq. (\ref{dif51})
\begin{eqnarray} &&
\partial_t E = -2s\pi K u V \,,
\label{dif13} \end{eqnarray}
On the other hand, we can use the equation (\ref{dif1}) to obtain
\begin{eqnarray} &&
\partial_t E =-\frac{K^2}{\gamma }
\int {\rm d}^2 r\, (\nabla^2\varphi)^2 \,.
\label{dif14} \end{eqnarray}
Substituting here $\nabla^2\varphi$ by $2p\partial_x\varphi$ in
accordance with Eq. (\ref{dif2}) we get
\begin{eqnarray} &&
\partial_t E =-\gamma  V^2
\int {\rm d}^2 r\, (\partial_x\varphi)^2 \,.
\nonumber \end{eqnarray}
The main logarithmic contribution to the integral comes from the
region $a<r<p^{-1}$ where $\partial_x\varphi\approx -sy/r^2$. Thus
we obtain
\begin{eqnarray} &&
\partial_t E =-\pi s^2 \gamma V^2 \ln\left(\frac{1}{pa}\right)\,.
\label{dif15} \end{eqnarray}
Comparing the expression with Eq. (\ref{dif13}) we find the same
answer (\ref{dif16}).

\subsection{Extremely small $\varGamma$}
\label{ap:extreme}

Here we consider the flow velocity, excited by the moving
disclination in the case of the extremely small $\varGamma$. The
velocity is zero in the zero in $\varGamma$ approximation (the case
is considered in Appendix \ref{ap:logre}), therefore we examine the
next, first, approximation in $\varGamma$. We use the same
formalism and the same designations as in Appendix \ref{ap:expli}.

In accordance with Appendix \ref{ap:expli} solutions of the complete
set of nonlinear stationary equations can be represented in the
following form:
\begin{eqnarray} &&
\partial_x\tilde\varphi=
\partial_y B \,, \qquad
\partial_y\tilde\varphi=
-(H+\partial_x B)\,,
\label{dphi} \end{eqnarray}
\begin{eqnarray} &&
\mathrm{curl}\ \bm v =\frac{K}{2\eta}
\left[-\partial_y H+2u\partial_y B
+2us\partial_y\ln r +\Phi' \right]\,,
\label{drot} \\ &&
v_x =-\frac{K}{2\eta}\partial_y\nabla^{-2}
\left[-\partial_y H+2u\partial_y B
+2us\partial_y\ln r +\Phi' \right] \,,
\label{dvx} \\ &&
v_y =\frac{K}{2\eta}\partial_x\nabla^{-2}
\left[-\partial_y H+2u\partial_y B
+2us\partial_y\ln r +\Phi' \right] \,,
\label{dvy} \end{eqnarray}
where $B$, $H$ and $\Phi'$ are to be found from the equations
\begin{align} &&&
-\partial_y H + 2p\partial_y B
+\frac{\varGamma}{4}\nabla^{-2}(\nabla^2-2u\partial_x)
\left[-\partial_y H+2u\partial_y B+2us\partial_y\ln r
+\Phi' \right]=
\nonumber \\ &&&
-\frac{\varGamma}{2}\Bigl(\partial_y\nabla^{-2}
\left[-\partial_y H+2u\partial_y B+2us\partial_y\ln r
+\Phi' \right]\partial_y B+\partial_x\nabla^{-2}
\left[-\partial_y H+2u\partial_y B+2us\partial_y\ln r
+\Phi' \right](\partial_x B+H)\Bigr) \,,
\label{dtphi} \end{align}
\begin{eqnarray} &&
\Phi'=2\nabla^{-2}
[(\partial_x B+H)\partial_x\partial_y H
+ \partial_y B\partial^2_y H] \,,
\label{dPPhi} \\ &&
\nabla^2 B+\partial_x H=-2\pi s \delta(\bm r) \,.
\label{dsource} \end{eqnarray}

If $\varGamma$ is extremely small, $s^2\varGamma\ln^2(ua)\ll1$, the
solution of Eqs. (\ref{dtphi}-\ref{dsource}) can be continued to the
vicinity of the core. In the main approximation the solution for
$\varphi$ coincides with the solution for the angle $\tilde\varphi_L$
in the absence of the liquid motion. This case, examined in
\cite{RK91,BP88,PR90,DE96}, is described in Appendix \ref{ap:logre}.
The functions $B_L$ and $H_L$, corresponding to $\tilde\varphi_L$, are
given by
\begin{eqnarray}
\label{dL} &&
2pB_L=H_L=2spK_0(pr)\exp(-px)\,.
\end{eqnarray}
This solution results in
\begin{eqnarray}
\label{dPhiln} &&
\Phi'=2 s^2 p \frac{y}{r^2}
\ln\left(\frac{\mathrm{min}\{r,p^{-1}\}}{a}\right)\,.
\end{eqnarray}
Then, after neglecting the nonlinear right-hand side of Eq.
(\ref{dtphi}), we can find
\begin{eqnarray}
\label{dH} &&
H(\bm r)=\frac{4\pi s}{1+\varGamma/4}
\int\frac{\mbox d^2q}{(2\pi)^2}\exp(i\bm{qr})
\frac{pq^2-(s\varGamma p/4)(q^2+2iuq_x)
\ln\left({\mathrm{min}\{(qa)^{-1},\ (pa)^{-1}\}}\right)
}{(q^2-2ik_1q_x)(q^2+2ik_2q_x)}\,.
\end{eqnarray}
$B(\bm r)$ can be found in the similar way. Using $B$ and $H$,
from Eqs. (\ref{dvx},\ref{dvy}) we calculate such flow velocity
$\bm v(\bm r)$ that vanishes at infinity.

For $r\gg p^{-1}$ this solution coincides with expressions
(\ref{BC},\ref{circ1},\ref{flux1}) with
\begin{eqnarray} \nonumber &&
\zeta_1(0)=2s+\frac{2s^2p}{u}\ln\left(\frac{1}{pa}\right)\,.
\end{eqnarray}
For $pr\ll 1$ the expression (\ref{dH}) is reduced to (\ref{dL}) and
this region produces the main contribution to $\Phi'$ (\ref{dPhiln}).
The following expressions are obtained in the inner region ($pr\ll 1$)
from the solution (\ref{dphi}-\ref{dH}):
\begin{eqnarray} &&
\varphi_1 =\left(u-sp\ln\frac{1}{pa}\right)y+
spy\ln\frac{r}{a}\,,
\label{dphi1ln} \\ &&
\mathrm{curl}\ \bm v =\frac{Ks^2 p}{\eta}
\ln\frac{r}{a}\ \frac{y}{r^2}\,.
\label{drotln} \end{eqnarray}
A relation between $p$ and $u$ is fixed by the condition (\ref{dif9}),
leading to $u=sp\ln[1/(pa)]$, which is equivalent to Eq.
(\ref{dif16}). The flow velocity at $pr\ll 1$ and $\ln(r/a)\gg 1$ is
\begin{eqnarray}
v_\alpha=-\frac{s^2\varGamma}{8}V \epsilon_{\alpha\beta}
\nabla_\beta\left[y\ln^2(r/a)\right] \,,
\label{bc4} \end{eqnarray}
which corresponds to the stream function
\begin{eqnarray} &&
\Omega=-Vy-\frac{Ks^2 p}{4\eta}
y\ln^2\left(\frac{r}{a}\right) \,.
\label{appcomega1} \end{eqnarray}
The expansion over $\varGamma$ near the disclination is regular and
can be derived from Eqs. (\ref{rg4},\ref{rg5}) with the condition
$\nabla\varphi_1(a)\sim p$\,: $\tilde\varphi_L+uy$ is the zero term of
the series for $\varphi$, and expression (\ref{appcomega1}) represents
the zero and the first terms for $\Omega$.

Note that in accordance with Eq. (\ref{bc4}) in the limit
$\varGamma\to0$ the flow velocity tends to zero near the disclination
core, $v(a)/V=O(\varGamma)$, despite the fact that the disclination
itself moves with the finite velocity $V$, thus there is a slipping on
the disclination core in this limit.

\subsection{Solution with the complete order parameter}
\label{ap:core}

Here we consider the dynamic equations for the coupled velocity
field $\bm v$ and the complete order parameter
$\Psi=Q\exp(i\varphi/|s|)$ of the hexatic or smectic-$C$ films.
These equations are needed to examine the velocity field in a
vicinity of the disclination core. Below, we work in the framework
of the mean field theory.

Formally the equations can be derived, using the Poisson brackets
method \cite{KL93,DV80}. The energy, associated with the order
parameter, in the mean field approximation is
\begin{eqnarray} &&
{\cal H}_\Psi=\frac{Ks^2}{2}\int \mbox d^2r
\left(|\nabla\Psi|^2+\frac{1}{2a_s^2}\left(1-|\Psi|^2\right)^2\right)\,,
\nonumber
\end{eqnarray}
its density passes to the $K$-contribution in Eq. (\ref{aa1}) on
large scales $r\gg a_s$. The only non-trivial Poisson bracket,
which has to be added to the standard expressions (see
\cite{KL93,DV80}), is \cite{gur}
\begin{eqnarray} &&
\{j_\alpha(r_1),\Psi(r_2)\} =-\nabla_\alpha\Psi\delta(r_1-r_2)
+\frac{i}{2|s|}\Psi(r_2)\epsilon_{\alpha\beta}\nabla_\beta\delta(r_1-r_2) \,.
\nonumber
\end{eqnarray}
The expression for the energy and the Poisson bracket is correct
for hexatics, and turns to the above expressions at large scales
for the smectics-$C$ (due to fluctuations). Then the dynamic
equations read:
\begin{align}
\nonumber
\rho\partial_t v_\alpha + \rho v_\beta\nabla_\beta v_\alpha&=
\eta\nabla^2 v_\alpha-\frac{s^2 K}{2}
\left\{\nabla_\alpha\Psi^* \left(\nabla^2\Psi
+\frac{1}{a_s^2}\Psi\left(1-|\Psi|^2\right) \right)
+\nabla_\alpha\Psi \left(\nabla^2\Psi^*
+\frac{1}{a_s^2}\Psi^*\left(1-|\Psi|^2\right)\right) \right\}
\\
&-\frac{i|s| K}{4}\epsilon_{\alpha\beta}\nabla_\beta
\left\{\Psi^*\nabla^2\Psi -\Psi\nabla^2\Psi^*\right\}
+\nabla_\alpha\tilde\varsigma\,,
\nonumber\\
\partial_t\Psi+v_\alpha\nabla_\alpha\Psi&=
\frac{i}{2|s|}\Psi\epsilon_{\alpha\beta}\nabla_\alpha v_\beta
+\frac{Ks^2}{2\gamma_s}\left(\nabla^2\Psi+
\frac{1}{a_s^2}\Psi\left(1-|\Psi|^2\right)\right) \,,
\label{psiv}
\end{align}
the relation $\gamma_s=s^2\gamma/2$ ensures the reduction to Eq.
(\ref{IIi2}) in the limit $|\Psi|=1$, and the kinetic coefficients
are believed to be independent of $Q$ (otherwise one can assume,
for example, the dependence $\gamma_s=s^2\gamma|\Psi|^2/2$). The
slow dynamics of a $2d$ liquid crystalline system with
disclinations can be described by Eqs. (\ref{psiv}) with the
additional condition of incompressibility $\nabla\bm v =0$ which
enables to exclude the passive variable $\tilde\varsigma$.

If the distance from the disclination point to another disclination
is much larger than $a_s$, i.e. the perturbation to the static
solution $\Psi_0=Q_0\exp(i\varphi_0/|s|)$ for the single defect is
small, we can linearize Eqs. (\ref{psiv}) with respect to the
perturbation expressed in terms of corrections $Q_1$ and
$\varphi_1$ to $Q_0$ and $\varphi_0$ respectively:
\begin{align}
\label{vmodphasem}
&\eta\nabla^2 v_\alpha
-2\gamma_s \left\{\nabla_\alpha
Q_0(v_\beta-V_\beta)\nabla_\beta Q_0
+\frac{1}{s^2}Q_0^2\nabla_\alpha\varphi_0
\left((v_\beta-V_\beta)\nabla_\beta \varphi_0
-\frac{1}{2}\epsilon_{\beta\gamma}\nabla_\beta
v_\gamma\right)\right\}
\\ &
+2\gamma_s\frac{1}{2s^2}\epsilon_{\alpha\beta}\nabla_\beta
\left[Q_0^2\left((v_\mu-V_\mu)\nabla_\mu
\varphi_0-\frac{1}{2}\epsilon_{\mu\nu}\nabla_\mu v_\nu
\right)\right]+\nabla_\alpha\tilde\varsigma=0\,,
\nonumber
\end{align}
\begin{eqnarray} &&
\frac{Ks^2}{2\gamma_s}\left(\nabla^2 Q_1
-\frac{(\nabla\varphi_0)^2}{s^2}Q_1
-\frac{2\nabla_\alpha\varphi_1\nabla_\alpha\varphi_0}{s^2}Q_0
+\frac{1}{a_s^2}\left(1-3Q_0^2\right)Q_1\right)
=(v_\beta-V_\beta)\nabla_\beta Q_0 \,,
\label{Q1}
\end{eqnarray}
\begin{eqnarray} &&
\frac{Ks^2}{2\gamma_s}\left(\nabla^2\varphi_1
+2Q_0^{-1}(\nabla_\alpha Q_1 \nabla_\alpha\varphi_0
+\nabla_\alpha Q_0 \nabla_\alpha\varphi_1)
\vphantom{\frac{1}{1}}\right)=
-\frac{1}{2}\epsilon_{\alpha\beta}\nabla_\alpha v_\beta
+(v_\beta-V_\beta)\nabla_\beta \varphi_0 \,.
\label{p1}
\end{eqnarray}

In terms of dimensionless quantities $L=\eta\Omega/K$, $R=r/a_s$
and $\varGamma=2\gamma_s/(s^2\eta)$, Eq. (\ref{vmodphasem}) takes
the following form (as previously, we consider a disclination with
the unitary topological charge $|s|$ or $-|s|$):
\begin{eqnarray} &&
\nabla^4_R L + \frac{\varGamma}{4}\left\{
-4s^2 \frac{(\partial_R Q_0)^2}{R^2}\partial^2_\phi L +
\left(\nabla^2_R+\frac{2s}{R}\partial_R \right)
\left(Q^2_0 \left(\nabla^2_R-\frac{2s}{R}\partial_R\right)L\right)
\right\}=0\,,
\label{LL}
\end{eqnarray}
$Q_0$ is found from
\begin{eqnarray} &&
\nonumber
\left(\partial^2_R +\frac{1}{R}\partial_R-\frac{1}{R^2}\right) Q_0
+ Q_0(1-Q^2_0)=0\,,\qquad Q_0(0)=0\,,\qquad
Q_0(\infty)=1\,.
\end{eqnarray}
If $\varGamma\gg 1$, as it follows from Eq. (\ref{LL}), a new scale
$R\sim 1/\sqrt{\varGamma}\ll 1$ appears inside the core, the first
term in Eq. (\ref{LL}) can be neglected on larger scales and there
is no crossover at $R\sim 1$.

If $Q_0\equiv 1$,  Eq. (\ref{LL}) is reduced to Eq. (\ref{rg17}).
If $R\ll 1$, $Q_0=AR$ ($A\approx 0.58$), and Eq. (\ref{LL}) can
be rewritten as
\begin{eqnarray} &&
\nabla^2_R\left\{
\nabla^2_R L +
\frac{A\varGamma}{4}(R^2\nabla^2_R-4s^2)L\right\}=0\,.
\nonumber
\end{eqnarray}
The solution of the equation is a composition of terms
$\lambda(R)\sin(m\phi)$ with different $m$. The condition
$\lambda(R)=0$ keeps two constants in the general solution of the
ordinary differential equation for $\lambda(R)$, two regular near
$R=0$ partial solutions are
\begin{eqnarray} &&
R^{|m|} \quad
\mathrm{ and} \quad
R^{|m|}\,_2F_1\left(\frac{|m|-\sqrt{m^2+4s^2}}{2},\,
\frac{|m|+\sqrt{m^2+4s^2}}{2},\,1+|m|,\,
-\frac{A^2\varGamma R^2}{4}\right)\,,
\nonumber
\end{eqnarray}
where $_2F_1$ is the hyper-geometric function
($_2F_1(a,b,c,z)=1+abz/c+\dots$). Two constants (e.g. derivatives
$\lambda^{(|m|)}(0)$ and $\lambda^{(|m|+2)}(0)$) are chosen to
ensure the slowest possible grow at $R\gg 1$, i.e to exclude the
largest exponent among $\alpha$ in Eq. (\ref{rg19}).

If $\varGamma\gg 1$, it is possible to derive a better
approximation in the core. We can expand $Q_0(R)$
in a series, look for the series solution $\lambda(R)$
and extract the terms of the highest order in $\varGamma$.
For example, for $m=1$ the series for
$\lambda(R)$ begins with $l_1 R+l_3 R^3$, which fix two constants
in the partial solution:
\begin{align}
\lambda(R)=&l_1 R \left[1+\frac{1}{A^2\varGamma s (2-s^2)}
\left(-1-\frac{s^2 A^2\varGamma R^2}{8}
+ \,_2F_1\left(\frac{1-\sqrt{1+4s^2}}{2},\,
\frac{1+\sqrt{1+4s^2}}{2},\,2,\,
-\frac{A^2\varGamma R^2}{4}\right)\right)\right]
\nonumber \\
+&l_3\frac{8}{A^2 \varGamma s^2}R
\left[-1+ \,_2F_1\left(\frac{1-\sqrt{1+4s^2}}{2},\,
\frac{1+\sqrt{1+4s^2}}{2},\,2,\,
-\frac{A^2\varGamma R^2}{4}\right)\right]\,.
\nonumber
\end{align}
The solutions of Eqs. (\ref{Q1},\ref{p1}) have the following form:
\begin{eqnarray}
Q_1=\vartheta(R)\partial_\phi\sin(m\phi)\,,\qquad
\varphi_1=\sigma(R)\sin(m\phi)\,,
\nonumber
\end{eqnarray}
and should be found from the equations
\begin{eqnarray}&&
\vartheta''+\frac{1}{R}\vartheta'-\frac{1+m^2}{R^2}\vartheta
-\frac{2Q_0}{sR^2}\sigma+(1-3Q^2_0)\vartheta=
\varGamma\frac{1}{R}\partial_R Q_0\lambda\,,
\nonumber\\ &&
\sigma''+\frac{1}{R}\sigma'-\frac{m^2}{R^2}\sigma
+\frac{2}{Q_0}\left(-\frac{sm^2}{R^2}\vartheta
+\partial_R Q_0\sigma'\right)=
\frac{\varGamma}{2}\left(
\lambda''+\frac{1-2s}{R}\lambda'-\frac{m^2}{R^2}\lambda\right)\,,
\nonumber
\end{eqnarray}
which generalize the expressions of Ref. \cite{PR90}.

The dynamic equations with the complex order parameter demonstrate
that at any $\varGamma$ nothing special happens on the core in the
boundary conditions for Eqs. (\ref{lli1}-\ref{IIi2}). The
peculiarity of extremely small $\varGamma$ which leads to
non-slipping condition is in the slow grow of $\nabla\Omega$ far
from the disclination.

\end{document}